# Defect mediated changes in structural, optical and photoluminescence properties of Ni substituted CeO$_2$

Saurabh Tiwari[1], Gyanendra Rathore[2], N. Patra[3], A. K. Yadav[3], Dibyendu Bhattacharya[3], S.N. Jha[3], C.M. Tseng[4], S.W. Liu[2], Sajal Biring[2]*, Somaditya Sen[1,2]*

[1]Metallurgy Engineering and Material Sciences, Indian Institute of Technology Indore, Simrol Campus, Khandwa Road, Indore, India

[2]Electronic Engg., Ming Chi University of Technology, New Taipei City, Taiwan

[3]Atomic & Molecular Physics Division Bhabha Atomic Research Centre, Trombay, Mumbai, India

[4]Materials Engg., Ming Chi University of Technology, New Taipei City, Taiwan

**Abstract:**

Local and long range structure, optical and photoluminescence properties of sol-gel synthesized Ce$_{1-x}$Ni$_x$O$_2$ (0 ≤ x ≤ 0.1) nanostructures have been studied. The crystal structure, lattice strain and crystallite size have been analyzed. A decrease in lattice parameter may be attributed to substitution of Ce with smaller Ni ion. UV-Vis measurement is used for studying the effect of Ni substitution on bandgap and disorder. The bandgap decreases with Ni substitution and disorder increases. The PL spectra show five major peaks attributed to various defect states. The PL emission decreases with Ni substitution owing to increase in defects which acts as emission quenching centers. The lattice disorder and defects have been studied using Raman spectroscopy. Raman measurement shows that oxygen vacancies related defects are increasing with Ni substitution which causes changes in optical and PL properties. Local structure measurements show that Ni substitution leads to oxygen vacancies which does change host lattice structure notably. Ce$^{4+}$ → Ce$^{3+}$ conversion increases with Ni substitution.

Keywords: Cerium oxide; Nanoparticles; Strain; Bandgap; Urbach energy.

**Introduction:**

Energy crisis in the time of rapid development and increasing population is a major issue to be addressed. Sustainable energy sources needs be deployed for meeting present and future demands. Solar energy is one of the most important sources of sustainable energy. Photocatalysis, hydrogen production, dye-sensitized solar cells etc. are the different means to utilize solar energy [1–4]. Photocatalytic activity of various nanostructures on water splitting and degradation of water pollutants (dyes), have been recently investigated in details [5–9]. Rare earth oxides have attracted special attention. This is due to the peculiarity of 4f electrons, and CeO$_2$'s unique structural and optical properties. Cerium is most abundant material in earth crust [10]. Along with photocatalysis $CeO_2$ has applications in biomedical field due to being a reactive oxygen species (ROS) [11]. Natural sunlight spectra have only 2-3% of ultra violet (UV) light, while, ~ 45% is visible light [12]. $CeO_2$'s high bandgap ~ 3.1 eV is a problem as it cannot absorb the visible light but can only use the UV radiation, i.e. only 2-3% of the solar radiation. However, tailoring of bandgap may allow $CeO_2$ to absorb visible light. This is essential for its compatibility as solar applications. Doping, lattice defects, strain generation, etc. may create defects in the bandgap and thereby tune the bandgap [13–16]. Transition element doping in $CeO_2$ is a good example, which has been proved to greatly increase its catalytic and biomedical properties owing to reduction in bandgap and oxygen vacancies related defects [2,17]. Ni doping in $CeO_2$ reduces the bandgap from UV region to visible region [11]. It also increases in ferromagnetic behavior and catalytic properties [11,18]. Fluorite structured $CeO_2$ due to increased oxygen vacancies displays better optical and ferromagnetic properties [19,20]. This makes the materials more suitable for biomedical and photocatalytic applications with enhanced functionality [17,21,22].

A detailed analysis of defect mediated changes on structural, optical and photoluminescence properties of Ni substituted $CeO_2$ is being reported in this work for the first time. Samples were synthesized using sol-gel technique to ensure proper homogeneity of the samples. A detailed correlation between structural, optical and photoluminescence studies is established in this work.

**Experimental Procedure**

Pure and Ni-substituted $CeO_2$, $Ce_{1-x}Ni_xO_2$, nanoparticles were synthesized by sol-gel technique using cerium nitrate and nickel nitrate as precursors. Cerium nitrate and Ni nitrate (Alfa Aesar, purity 99.99%) were used as precursors. The $Ce_{1-x}Ni_xO_2$ nanoparticles (x=0, 0.025, 0.05 and 0.1) were synthesized and referred as CNO0, CNO2, CNO5 and CNO10 respectively. Detailed synthesis procedure is already reported [23].The resultant powders were further calcined at 450°C for 6h to release trapped carbon particles.

High Resolution Transmission Electron Microscope (HRTEM) images, Selected Area Electron Diffraction (SAED) and Energy Dispersive X-ray spectroscopy (EDX) were performed with HRTEM JEOL JEM-2100 $LaB_6$, operated at voltage of 200kV. For XRD, a Bruker D2-Phaser x-ray Diffractometer is used for structure analysis. UV-vis spectroscopy was performed using a Shimadzu UV-2600 spectrophotometer. Raman spectra were taken with a HORIBA-Scientific spectrometer with excitation wavelength 632.8nm. X-ray Absorption Spectroscopy (XAS) measurement have been carried out on Ni doped $CeO_2$ at Ni K-edges and Ce $L_{III}$ edge to probe the local structure of the samples. XAS comprises of both X-ray Near Edge Structure (XANES) and Extended X-ray Absorption Fine Structure (EXAFS) spectroscopy. The XAS measurements have been carried out at the Energy-Scanning EXAFS beamline (BL-9) at the Indus-2 Synchrotron Source (2.5 GeV, 200 mA) at Raja Raman Centre for Advanced Technology (RRCAT), Indore, India. This beamline operates in the energy range of 4 KeV to 25 KeV. The details and mechanism has been reported earlier [24].

**Results and discussion**

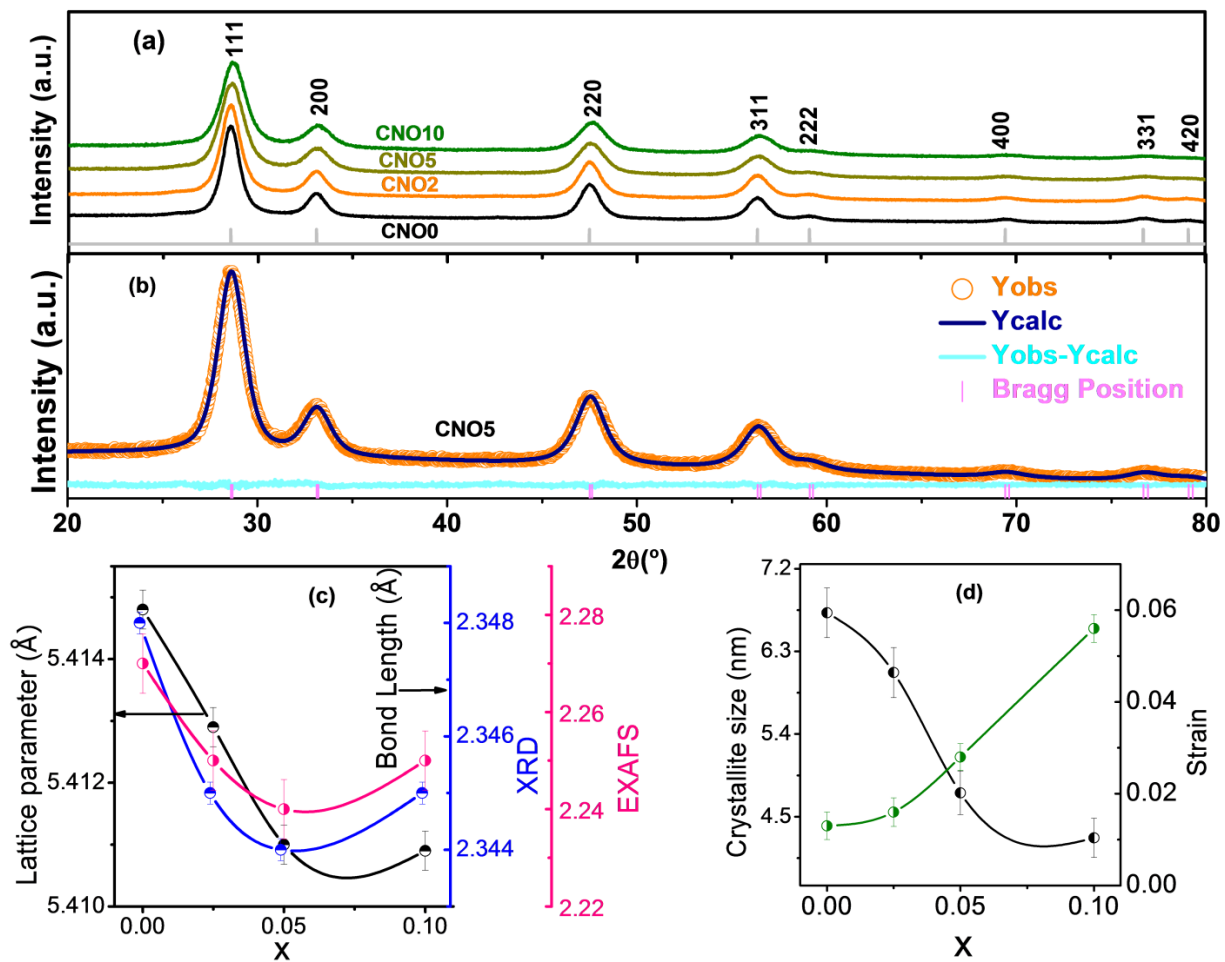

Fig. 1. (a) XRD data of $Ce_{1-x}Ni_xO_2$ samples confirms a cubic fluorite structure having space group Fm3m (-inset shows reduction of crystallite size with x) (b) Variation of lattice parameter and strain with x.

XRD patterns of synthesized CNO0, CNO2, CNO5 and CNO10 (Figure 1(a)) shows (111), (200), (220), (311), (222), (400), (331) and (420) reflection planes of a cubic fluorite structure with space group Fm3m. No impurity phases of oxides of Ni were found. This hints at proper substitution of Ni in the $CeO_2$ lattice. Ni, in whatever valence state, is smaller in size than $Ce^{4+}$(VIII) (1.11 Å). Rietveld refinement was carrying out by FullProf software to evaluate changes in lattice parameter due to Ni substitution (Figure 1(b)). Lattice parameters reduce with substitution (Figure 1(c)). Hence, unit cell volume is reduced. Since the size of Ni ions is smaller in comparison to Ce, may be a reason behind lattice contraction. Debye-Scherer's equation: $D= K\lambda/\beta cos\theta$ (where, K=0.9) is used for calculation of crystallite size. Crystallite size decreases with Ni substitution from 6.7 to 4.3 nm (Figure 1(d). Strain was calculated using Williamson-Hall equation [23]: $[\beta Cos\theta/\lambda = 1/D + €Sin\theta/\lambda]$, where β is full width at half maxima(FWHM), D is crystallite size, € is lattice strain, θ is angle of incidence and λ is wavelength (1.5406 Å) of Cukα radiation. Strain increases with substitution (Figure 1(d)). The decrease in crystallite size with Ni doping may be due to increased strain and defects which impede long range ordering.

To further confirm the microstructure and particle size, HRTEM study is performed on the samples. Aggregation of crystallites is observed which is common in nanoparticles synthesized by wet chemical method. Nanoparticles achieve more stable energy state by agglomeration. Also the presence of hydroxyl ions facilitates agglomeration [25]. HRTEM images of CNO0 and CNO5 nanocrystals are shown in [Figure 2 (a, b)]. For CNO0, average size crystallite is ~7.4 (± 0.5) nm. With Ni substitution, size reduces to ~5 (± 0.5) nm for CNO5. These results follow XRD results. Nanoparticles appears to have thermodynamically stable (111) and (200) termination faces [26]. Well-defined lattice fringes in SAED patterns confirm the crystalline nature of synthesized samples [Figure 2(a, b insets)]. HRTEM studies are in agreement with XRD analysis. EDX studies reveals presence of Ce in both samples and incorporation of Ni in CNO5.

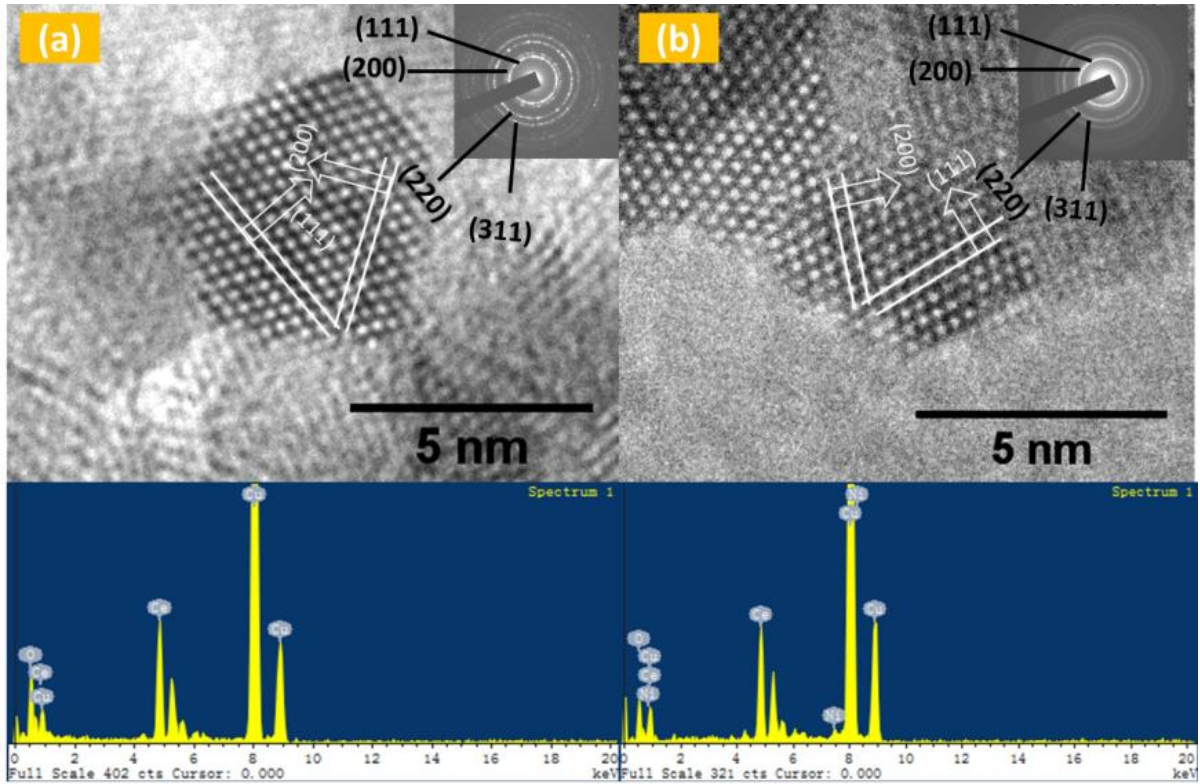

Fig. 2: HRTEM images revealing average crystallite size of (a) CNO0 ~7.4nm and (b) CNO5 ~5nm with (111) and (200) plane termination for both samples. (Insets) SAED ring patterns show (200), (111), (220) and (311) reflections with d-spacing corresponding to $CeO_2$ structure, (c) EDX patterns of CNO0 and CNO5 showing elemental components of the samples.

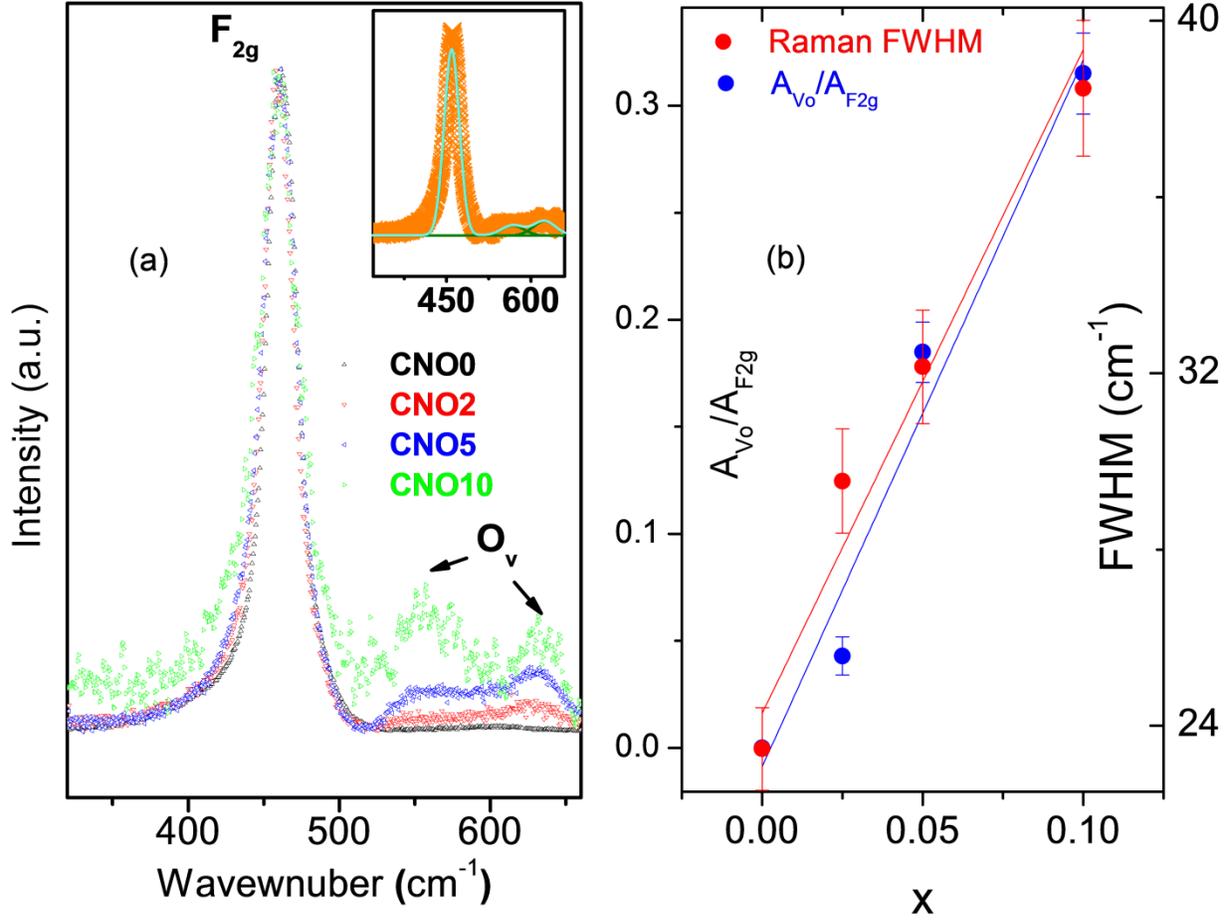

**Fig. 3.** (a) Systematic broadening and nominal red shift of $F_{2g}$ phonon mode at ~460 cm$^{-1}$ in modified $Ce_{1-x}Ni_xO_2$; a broad feature in the range 520-660 cm$^{-1}$ associated with oxygen defects becomes stronger with substitution (-inset shows Raman peak fitting) (b) shows increase in Vo fraction and FWHM with Ni doping in $CeO_2$.

All CNO samples show (Figure 3 (a)) a strong Raman peak at 462.7 cm$^{-1}$ and a broad feature ranging from ~520 cm$^{-1}$ to ~660 cm$^{-1}$. The strong peak ~462.7 cm$^{-1}$ attributes to an $F_{2g}$ triply degenerate mode of $CeO_2$ fluorite structure. This represents a $CeO_8$ symmetric stretching vibration. Hence, this mode is highly sensitive to oxygen related disorder [27]. This peak broadens and becomes more asymmetric (lower half widths are larger than higher half width) with substitution [Fig. 4(a, b)]. Several factors like defects, strain, phonon confinement, and size distribution can be responsible for these changes [28]. Size plays an important role in Raman peak shifts [29]. Most probably reduction of size is responsible for red shift. However, a detailed study on confinement and strain reveals that most probably Raman phonon modes in CNO materials are affected by quantum confinement effects as well as inhomogeneous strain [29]. In defect-free crystalline semiconductor (bulk) materials one expects perfect symmetry. Asymmetry increases with decreasing crystallite size and is most prominent when size is smaller than Bohr excitation radius ~7.2nm of $CeO_2$ [28,30,31]. Since with reduction of size and dimension an increasing volume of reciprocal space is involved which leads to participation of phonons of other than zone boundary [28]. Hence, presence of considerable asymmetry strongly correlates quantum confinement effect in CNO system.

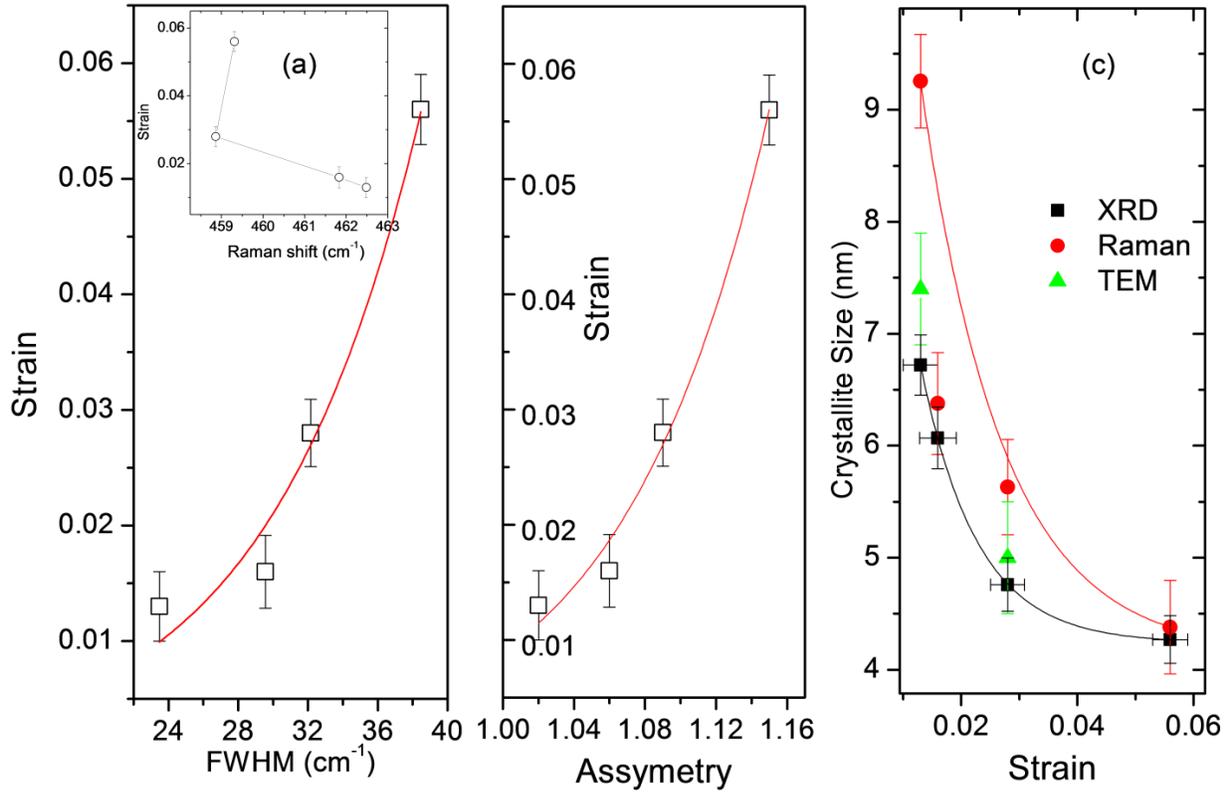

**Fig. 4.** (a, b) Strain variation with FWHM, peak position and asymmetry (b) Crystallite size variation with strain

A broad peak at ~520-660 cm$^{-1}$ corresponds to oxygen related defects [32,33]. This feature increases with substitution. Ni will be either in 2+ or 3+ valance state, it will cause generation of Vo to maintain charge neutrality of the lattice. A qualitative analysis of $V_O$ fraction has been estimated from $A_{Vo}/A_{F2g}$ ratio. This ratio was found to increase indicating that, with increased substitution increase oxygen vacancies increasing (Figure 3 (b)). This may lower the bandgap and quench the PL emissions. Increase in asymmetry and FWHM confirms increased strain as shown in [Figure 4 (a, b)]. Crystallite size was also estimated and is in unison with TEM and XRD results.

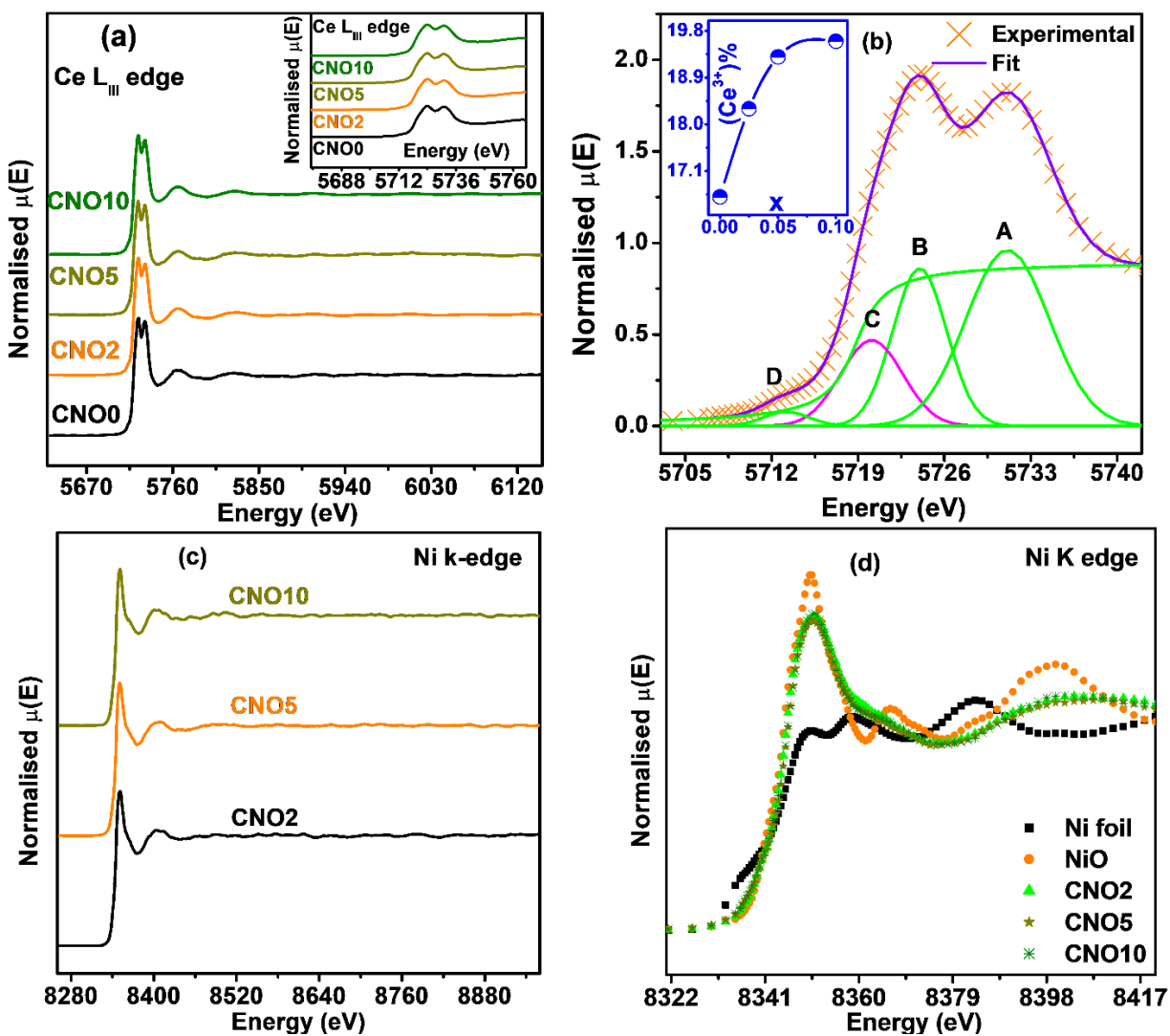

**Fig. 5.** (a, b) EXAFS spectra at Ce-$L_{III}$ edge and XANES spectra fitting of CNO0, CNO2, CNO5 and CNO10 respectively using Athena software shows $Ce^{3+}$ concentration increasing with Ni doping in $CeO_2$ (c, d) EXAFS spectra at Ni-K edges and XANES spectra; shows Ni is 2+ valance state.

Valence states of constituent elements were examined using XANES analysis of CNO samples. Ce $L_{III}$ edges of pure and Ni-substituted samples consist of two major peaks (Figure 5 (a)-inset). All samples have similar shapes, which were fitted by a combination of an arctangent and Gaussian functions (figure 5 (b). Four peaks, A, B, C and D are noted at 5730.9 eV, 5724.1 eV, 5719.3 eV and 5714.1 eV respectively (Figure 5 (b). A and B represents mixture of multi-electron transitions with final state of $2p4f^0 5d*$ and $2p4f^1 5d*L$. For above notations, $2p \rightarrow$ Ce $2p$ holes, $5d* \rightarrow$ excited electrons in $5d$ states and L $\rightarrow$ holes in anion ligand orbital (O $2p$). Peak C corresponds to Ce in $Ce^{3+}$ valence state. Whereas, Peak D can be attributed to final states of O$2p$-Ce$5d$ hybridizations due to crystal field splitting of Ce $5d$ states causing delocalization of d character at the bottom of conduction band [34]. The quantification of Ce valance in all samples has been done by using area ratio of $Ce^{3+}$ (C) and $Ce^{4+}$ (A+B) as [34–36]:

$$Ce^{3+} = \frac{C}{A+B+C} \times 100$$

$$Ce^{4+} = \frac{A+B}{A+B+C} \times 100$$

XANES analysis reveals a major $Ce^{4+}$ state (~84-80%) in pure and modified $CeO_2$ samples. However, $Ce^{3+}$ states are not negligible (~16-20%). Concentration of $Ce^{3+}$ increases in Ni-substituted samples from ~16% to ~20% for pure and Ni doped samples respectively [Figure 5 (b)]. Note that in $CeO_2$, Ce is expected to be primarily in $Ce^{4+}$ state. Hence, presence of $Ce^{3+}$ hints at O-vacancies, $V_O$, in lattice. On the other hand, Ni is predominantly in $Ni^{2+}$ state only [Figure 5(d)]. Nickel addition can induce, Vo due to lower valence state, $Ni^{2+}$. Hence, we expect a lot of oxygen defects in the Ni-incorporated CNO system, increasing with amount of substitution. $O^{2-}$ (IV) has a crystal radius of 1.24 Å, comparable to $Ce^{3+}$(VIII) (crystal radius ~1.28 Å). Both are larger than $Ce^{4+}$(VIII) (1.11 Å ). Similarly, $Ni^{2+}$(VIII) (~0.9 Å) is much smaller than all the three above. Lack of oxygen in the pure $CeO_2$ is a common feature and is observed using various types of synthesis [29]. $Ni^{2+}$ has a lesser valence state. This leads to further oxygen removal from lattice. Moreover, $Ni^{2+}$ is also smaller in size. Hence, the situation demands space and charge compensation by a larger in size and more charged ion. Hence, an increment in $Ce^{3+}$ concentration is logical.

The local structure of the absorbing atom is obtained from quantitative analysis of Ce $L_{III}$ edge and Ni K edge EXAFS spectra [Figure 5 (a, c)]. The absorption function $\chi$ (E) was obtained from absorption spectra $\mu$ (E) as follows [37]:

$$\chi(E) = \frac{\mu(E) - \mu_0(E_0)}{\Delta\mu_0(E_0)} \qquad (2)$$

where, $E_0$ is absorption edge energy, $\mu_0$ $(E_0)$ is the bare atom background and $\Delta\mu_0$ $(E_0)$ is the step size of $\mu(E)$ at the absorption edge. The wave number dependent absorption coefficient $\chi$ (k) was obtained from energy dependent absorption coefficient $\chi(E)$ as follows,

$$k^2 = \frac{2m(E-E_0)}{\hbar^2} \qquad (3)$$

where, m is the mass of electron, for amplification of oscillation at high $k$, $\chi$ (k) is weighted by $k^2$ and the $\chi$ (R) versus R spectra are generated by Fourier transform of $\chi(k)$ $k^2$ functions in R-space, in terms of the real distances from the center of the absorbing atom.

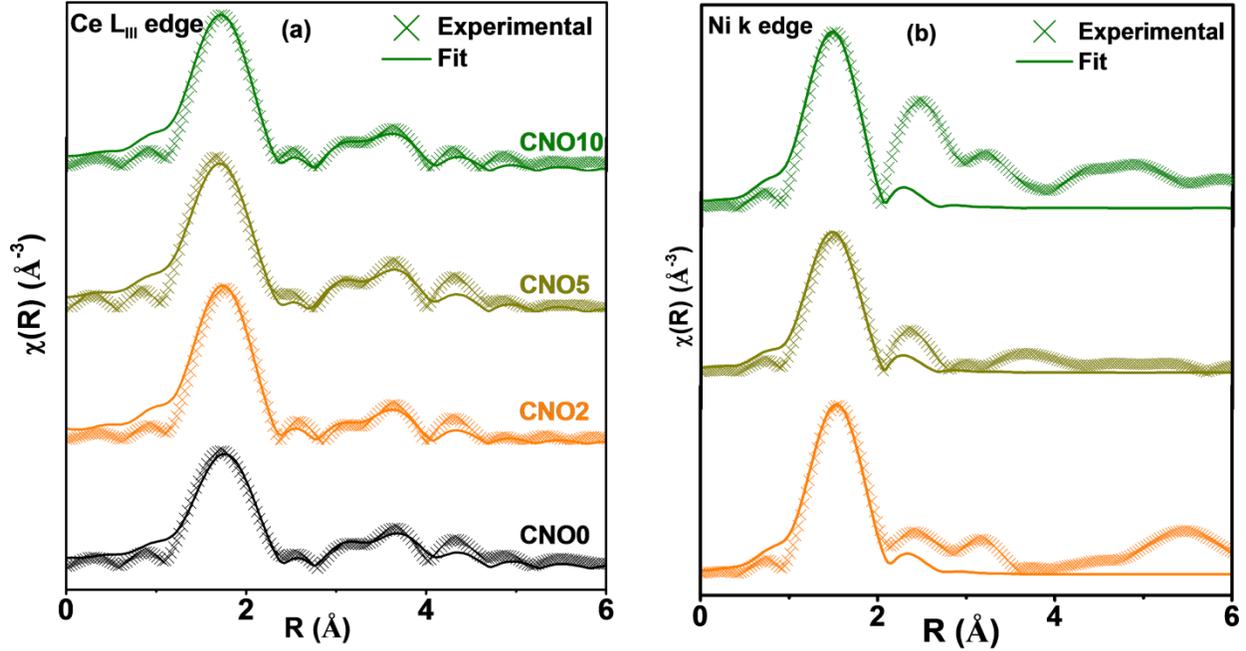

**Fig. 6.** (a, b) EXAFS spectra at Ce-$L_{III}$-edge and Ni K-edge respectively.

The EXAFS data analysis have been performed using set of EXAFS data analysis program available within Demeter software package [37]. It is having ATHENA software which is used for background reduction and Fourier transform to derive $\chi$ (R) versus R spectra from the absorption spectra and having ARTEMIS software which is used for generation of the theoretical EXAFS spectra starting from an assumed crystallographic structure and finally fitting of experimental data with the theoretical spectra. Theoretical fits are obtained by adopting a standard pure $CeO_2$ structure and thereafter refining bond length, coordination number and disorder factor. The theoretical and experimental data of Ce $L_{III}$ edge for all the samples shows good fits [Figure 6 (a, b)]. Since the spectrum range at Ni K edge is small, only first peak in the Fourier transform spectra is used for fitting. The obtained fitting results are shown in Table 1 and 2. In the Ce $L_{III}$ data, the first peak at 1.75 Å represents Ce-O coordination at 2.27 Å. The coordination peak between 3-4 Å is fitted with contribution of Ce and O atoms at 3.78 Å and 4.40 Å respectively. It can be observed from the fitting results in table 2 that the bond length is relatively larger compared to other Ni-O bond distances. This agrees with the XANES results.

The first peak in CNO samples has contributions of 8 oxygen atoms around Ce at ~2.29 Å (observed at a smaller bond length without phase correction). The observed bond length is a bit smaller than the Ce–O bond length (2.33 Å) in $CeO_2$. The second peak has contributions of the second coordination shell of 12 Ce atoms at ~ 3.5 Å and the third peak has contribution of the third coordination shell containing 24 oxygen atoms at ~ 4 Å. Oxygen coordination is found to be decreasing with doping, which hints at increase in oxygen vacancies with doping, confirmed by Raman, UV and PL measurements also.

Table 1: Bond length, coordination number and disorder factor obtain by EXAFS fitting for Ce L$_{III}$-edge.

| Path | Parameters | CNO0 | CNO2 | CNO5 | CNO10 |
|---|---|---|---|---|---|
| Ce-O (1) | R (Å) | 2.27 ± 0.01 | 2.25 ± 0.042 | 2.24 ± 0.045 | 2.25 ± 0.053 |
|  | N | 8 | 7.92 | 7.92 | 7.20 |
|  | $\sigma^2$ (10$^{-3}$ Å$^2$) | 8.7 ± 1.4 | 7.7 ± 0.9 | 10.1 ± 1.7 | 8.6 ± 1.4 |
| Ce-Ce | R (Å) | 3.78 ± 0.0081 | 3.77 ± 0.0137 | 3.76 ± 0.0174 | 3.77 ± 0.0113 |
|  | N | 12 | 11.88 | 11.88 | 10.80 |
|  | $\sigma^2$ (10$^{-3}$ Å$^2$) | 8.3 ± 0.7 | 12.9 ± 1.8 | 10.4 ± 1.3 | 10.9 ± 1.6 |
| Ce-O (2) | R (Å) | 4.40 ± 0.0073 | 4.39 ± 0.0128 | 4.39 ± 0.0192 | 4.38 ± 0.0171 |
|  | N | 24 | 23.76 | 23.76 | 21.60 |
|  | $\sigma^2$ (10$^{-3}$ Å$^2$) | 7.6 ± 1.2 | 10.7 ± 1.5 | 9.1 ± 1.1 | 11.7 ± 1.9 |

The Ce-O bond length decreases with doping as confirmed by XRD and EXAFS which leads to lattice distortion and strain. Increased strain and disorder as estimated from XRD and Urbach energy calculations confirmed by increased disorder factors of the Ce–O as well as Ce–Ce shells. The EXAFS signal of Ni K-edge data for Ni substituted samples exhibit a contribution like the Ce L$_{III}$-edge of the pure CeO$_2$ [Figure 6 (a, b)]. This analogous behavior promptly indicates the substitution of Ce by Ni. The first peak in Ni substituted CeO$_2$ has contribution of first oxygen shell at distance of 2.05 Å. In Ce L$_{III}$-edge EXAFS measurements, this bond length is observed to be somewhat lesser than Ce–O bond lengths. Reduction in Ce–O bond length is expected owing to higher ionic radii of Ce$^{4+}$ than the Ni$^{2+}$. One also observes a shorter Ce–O and Ce–Ce bond lengths for the Ni doped samples in comparison to the pure CeO$_2$. The change in interatomic spacing R$_{Ce-O}$ (first peak) is result of interaction between either Ce–O and Ce $-$ V$_o^{**}$ or between Ni-O and M$'_{Ce}-$V$_o^{**}$. The Ce-O-Ce/Ce-O-Ni interactions affects interatomic spacing R$_{(Ce-Ce/Ce-Ni)}$. Ohashi et al. reports alike results for Gd doped CeO$_2$ and reported that with increase in Gd-content, Ce–O and Gd–O bond length decreases [38].

Table 2: Bond length, coordination number and disorder factor obtain by EXAFS fitting for Ni K-edge.

| Path | Parameters | CNO2 | CNO5 | CNO10 |
|---|---|---|---|---|
| Ni-O | R (Å) | 2.05 ± 0.003 | 2.02 ± 0.007 | 2.02 ± 0.004 |
| | N | 5.25 | 5.40 | 6.0 |
| | $\sigma^2$ ($10^{-3}$ Å$^2$) | 3.4 ± 0.2 | 1.0 ± 0.07 | 2.5 ± 0.13 |

The decrease of both Gd–O and Ce–O distances are results of decrease in the inter-atomic distance caused by oxygen ions adjacent to oxygen vacancy around both Ce and Gd ions get relaxed towards to their nearby vacancies. The reduction in the coordination numbers hints oxygen deficiency neighboring Ce, this causes straining effect and decrease in R(Ce-O) as well as in R(Ce-Ce).

Bandgap of the materials was estimated from UV-vis spectroscopy [Figure 7 (a, b)]. $CeO_2$ shows both direct and indirect bandgap [39]. Extrapolation of *(αhν)$^2$* vs *hν* and (αhν)$^{1/2}$ vs hν plots, where, α is the absorption coefficient, helped to estimate the direct and indirect bandgaps, respectively. Absorption edge arises from direct transitions from top of the valence band (O 2p states) to empty 4f-shells of $Ce^{4+}$ [39]. Sharpness of the absorption edge in UV-vis spectra reduces and increasing amount of tailing (Urbach tail) is observed with substitution. Structural deformations and disorder are responsible for such tail formation. Urbach energy, $E_U$, was calculated using: *α = α$_{0*}$ exp(E/E$_U$)*, where *α* is absorption coefficient [40]. Calculation of α was done by taking natural logarithm of exponential decay (only tail part) near the absorption edge. Inverse of the slope of a straight line fit represents $E_U$ [-inset Figure 7 (c)]. Urbach energy increases with Ni doping from 0.153 eV for x=0 to 0.698 eV for x=0.1[Figure 7 (c)]. Increase in Urbach energy indicates increased defects and disorder in $CeO_2$ with Ni doping. These defect states not only cause of reduction of bandgap but also results in phonon assisted electronic transitions. Due to smaller crystal radius and lesser charge of Ni compared to Ce, Ni substitution in $CeO_2$ leads to lattice distortion and reduction of $Ce^{4+}$ to $Ce^{3+}$ owing to increased Vo concentration which highly modifies the band structure of $CeO_2$. Hence these changes lead to formation of localized states and spatial distribution of band structure which give rise to the possibility of both types of electronic transition direct and indirect.

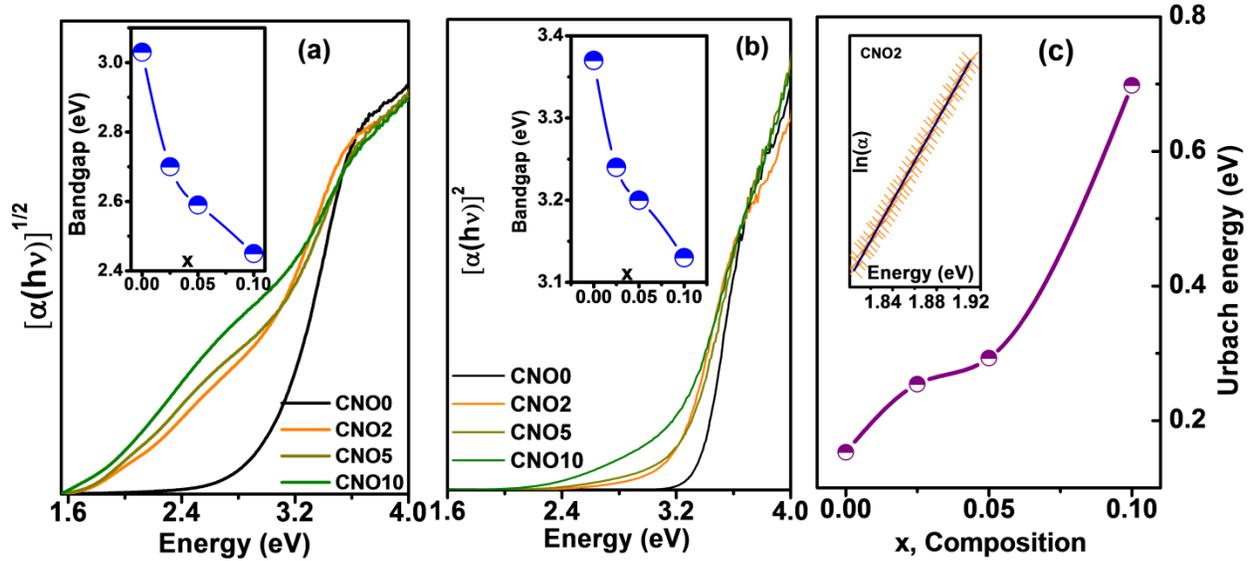

Fig. 7. Tauc-plot of $Ce_{1-x}Ni_xO_2$ for (a) Indirect transition (-inset shows bandgap decreasing with x from 3.03 to 2.45 eV), (b) Direct transition (-inset shows bandgap variation with x from, 3.37 to 3.13 eV) (c) Urbach energy increases with x from 0.153 to 0.698 eV (-inset shows Urbach fitting).

In an ideal situation where all Ce ions are in $Ce^{4+}$ state, the $CeO_2$ bandgap ~6eV is formed due to O2p (valence band, VB) and Ce 5d, 6s (conduction band, CB) [41]. A narrow sub-band ~1eV is formed due to the presence of empty $Ce^{4+}4f^0$ states [42]. This effectively reduces the bandgap to ~4.58eV (above VB) [41]. However, if $Ce^{3+}$ ions exist, a $Ce^{3+}4f^1$ sub-band is formed ~1.2-1.5eV above the VB and 3.3eV below the CB [42]. This band behaves as a hole trap in the bandgap. This creates an energy gap of ~3.1-3.3eV between available electrons in $Ce^{3+}4f^1$ and the empty $Ce^{4+}4f^0$ sub-band. On the other hand, presence of oxygen vacancies, $V_O$, may create states inside the bandgap. Generally, a $V_O$ can trap two electrons. The trapped state is generally represented in literature as F or $F^0$ state. The $F^0$ state may lose one electron and become $F^+$ state and further lose two electrons to become $F^{++}$ state [43]. Theoretical calculations show that $F^+$ states are closer to VB compared to $F^0$ states [44,45]. However, $F^{++}$ centers are located at about ~0.15-0.2eV below the CB [46]. $F^{++}$ states act as electron traps [46]. There is a fundamental difference between the $F^{++}$ state and the $F^+$ or $F^0$ states. While $F^{++}$ have no localized electrons $F^+$ and $F^0$ states have electrons and are partially and fully occupied states [43]. The $F^+$ and $F^0$ states are ~1.7eV and ~2.1eV above the VB and ~3eV and ~2.74eV below CB [46,47]. However, excited states of $F^+$ and $F^0$ states are also empty states and are located near the CB and are denoted as $F^{+*}$ and $F^{0*}$. $F^{+*}$ and $F^{0*}$ excited states are ~0.4eV below CB [46]. Difference between $F^0$ and $F^{0*}$ is ~2.1eV, while, that of $F^{+*}$ and $F^+$ is 2.4-2.5eV [46]. Various complex defects like $Ce^{3+}–V_O–Ce^{3+}$ may also exist [48]. Such complex defect states has been explained as accidental recombination sites of a hole due to $Ce^{3+}$ and an electron at the $V_O$ site [49]. Such defects show a Stokes-shifted broad luminescence [50]. Hence, such defects are located everywhere in the bandgap. Hence, it was found logical to fit a wide band PL data with multiple peaks to observe not only the major contributions involving well known defects states but also gather knowledge about how complex defects are distributed in the lattice. It is also to be noted that a $Ce^{3+}4f^1$ electron is not actually located at a single location but in fact can travel from one Ce atom to another. Thereby, depending on the availability of $V_O$ sites in the near vicinity and the distortions in the lattice the respective energy levels are supposed to behave as multiple defect possibility in the lattice, forming simple and complex defect states all along the bandgap.

Photoluminescence (PL) spectra of the samples reveal multiple peaks between 395-529 nm [Figure 8 (a)]. Multiple peaks have been fitted to analyze the contribution of each type of color contributions in the PL spectra. The major bands are obtained at ~424 nm (2.92 eV, violet), ~457 nm (2.71 eV, blue), ~485 nm (2.56 eV, green-blue) and ~528 nm (2.34 eV, green). Apart from the above major contributions, very feeble contributions of yellow (2.1-2.17eV; ~579nm), orange (2-2.1eV; ~590nm) and red (1.65-2eV; ~620nm) emissions also take place. But these emissions are negligible compared with the major bands.

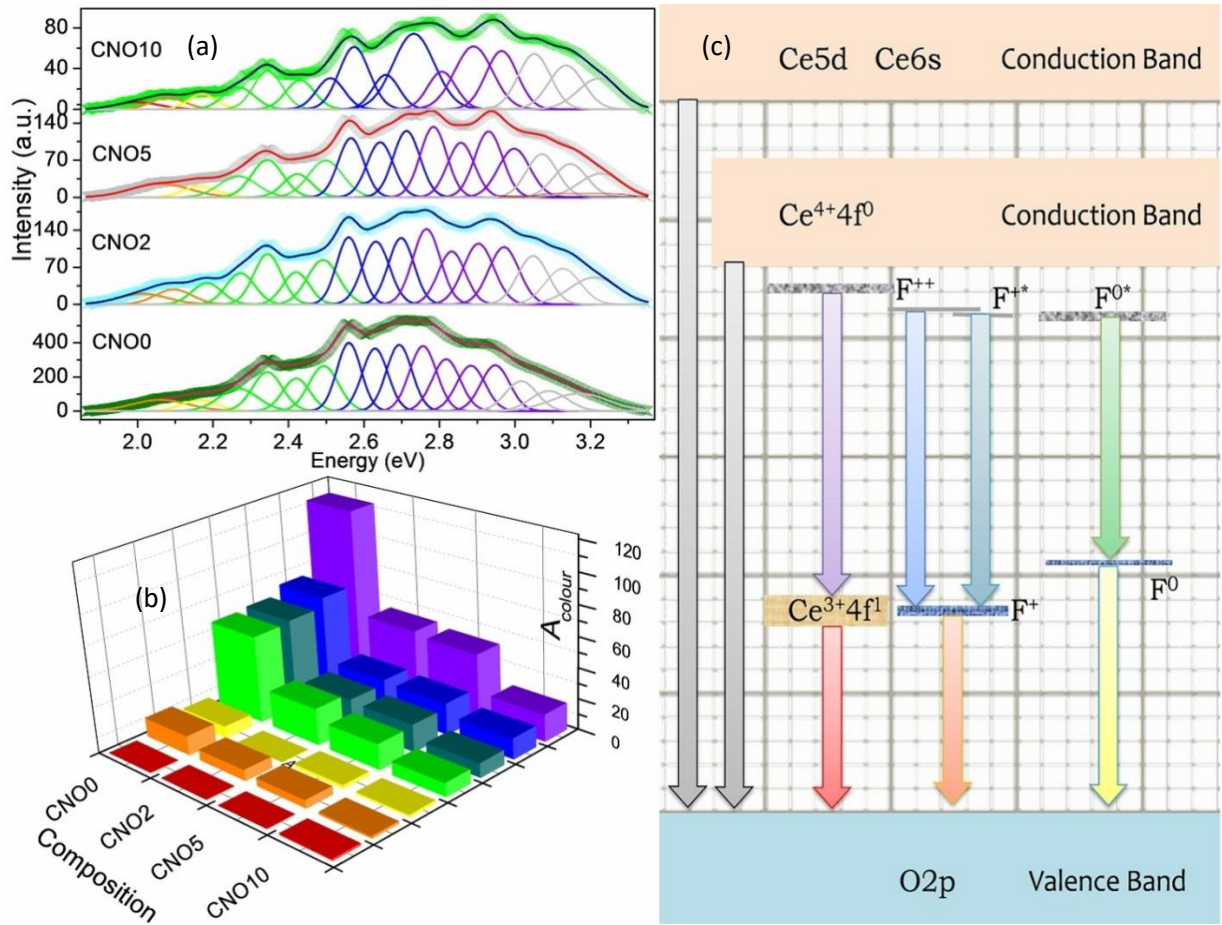

**Fig. 8.** (a) Photoluminescence spectra of $Ce_{1-x}Ni_xO_2$ samples (b) shows area variation of each color with composition (c) PL mechanism in Ni substituted $CeO_2$; shows various emission quenches with doping Ce-O bond length decreases with doping.

The major violet band may be ascribed to $F^{++} \rightarrow Ce^{3+} 4f^1$ transitions [51–59] . The blue-green and blue emissions are mostly a resultant of $F^{+*} \rightarrow F^+$ transitions while the green light comes from $F^{0*} \rightarrow F^0$ transitions [Figure 8 (c)]. PL intensity of the samples decreases with substitution. Individual color contribution decreases in a similar way [Figure 8 (b)].Multiple non-radiative decay processes may happen due to multi-phonon processes in the system. These non-radiative processes seem to enhance with substitution. The multi-phonon processes increases most probably due to distortion in the lattice. $Ni^{2+}$ incorporation increases $V_O$ and thereby converts $Ce^{4+} \rightarrow Ce^{3+}$ concentration at the surface and grain boundaries. These act as trap centers and behave as non-radiative recombination center, thereby repressing the emission intensity [43,60,61].

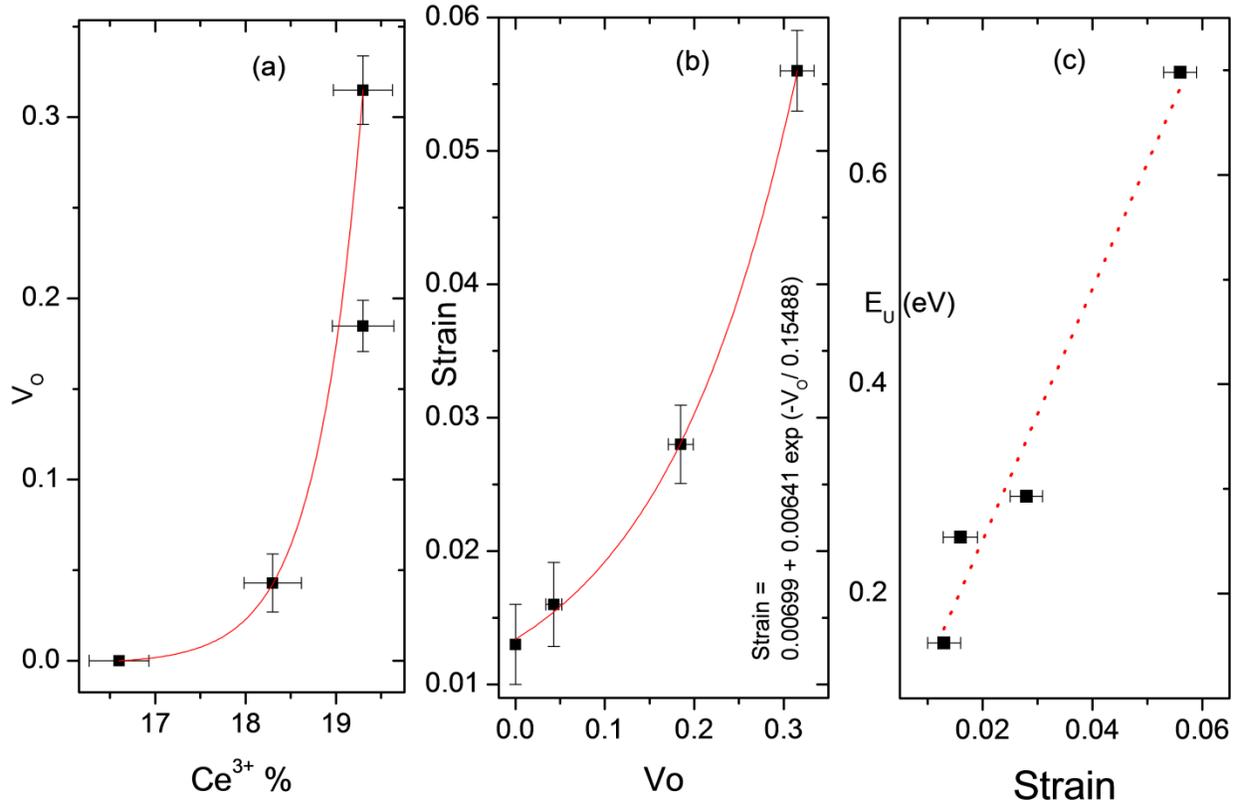

**Fig. 9.** (a) shows $Ce^{3+}$ vs Vo relationship with increase in $Ce^{3+}$ % Vo increasing exponentially (b) with increase in Vo strain increases in lattice following exponential relationship (c) Shows strain and Eu increasing linearly.

As $Ni^{2+}$ substitute $Ce^{4+}$ to maintain charge neutrality Vo increases in the lattice to maintain charge neutrality. The increasing $V_O$ concentration has two electrons which further leads to reduction of $Ce^{4+}$ to $Ce^{3+}$ [Figure 9 (a, b and c)] and also increases lattice strain. This reflects in electronic structure and thereby reduces bandgap.

## Conclusion

Single phase, homogeneous, nanocrystalline, sol-gel synthesized Ni substituted $CeO_2$ have the same cubic fluorite structure as that of pure $CeO_2$. XRD and HRTEM analysis confirm each other with the same crystal structure and crystallite size. Strain seems to increase while crystallite size decreases with substitution. On the other hand, the bandgap reduces a lot due to Ni substitution. As $Ni^{2+}$ substitute $Ce^{4+}$ to maintain charge neutrality Vo increases in the lattice to maintain charge neutrality. The increasing $V_O$ concentration has two electrons which further leads to reduction of $Ce^{4+}$ to $Ce^{3+}$ [Figure 9 (b)] and increases lattice strain. This reflects in electronic structure and thereby reduces bandgap. Reduction in bandgap is due to increased disorder mainly due to formation of defect states between valance band and conduction band. Raman study reveals that the $F_{2g}$ peaks become asymmetric and FWHM increases with doping. This implies increased strain in lattice. Also, $V_O$ related peaks increases with doping which hints increase in defect concentration with doping. Increased defect concentration may be due to difference ionic radii and valence states. EXAFS studies reveal reduction of oxygen coordination which implies increase in $V_O$ concentration with doping. Thus Ni substitution of Ce causes lattice distortion but it maintains the structure of $CeO_2$. XANES analysis also supports such observations as $Ce^{4+}$ is reduced to $Ce^{3+}$ as $Ce^{3+}$ concentration increases with doping. With increased substitution PL intensity of the samples decreases. The non-radiative processes increase with substitution. The multi-phonon processes increase with substitution due to increasing defect states and lattice distortion which leads to non-radiative emissions.

## Acknowledgements

Principle investigator expresses sincere thanks to Indian Institute of Technology Indore for funding the research. The authors sincerely thank Sophisticated Instrument Centre (IIT Indore) for FESEM studies, Mr. V. K. Jain (Amity University) for UV-vis analysis and Mr. Manoj Kumar (IISER Bhopal) for Raman studies. One of the authors (Dr. Sajal Biring) acknowledges the financial support from Ministry of Science and Technology, Taiwan (MOST 105-2218-E131-003) and 106-2221-E-131-027).